\documentclass[11pt]{article}

\usepackage[preprint]{acl}

\usepackage{times}
\usepackage{latexsym}

\usepackage[T1]{fontenc}

\usepackage[utf8]{inputenc}
\usepackage[table]{xcolor}
\usepackage{booktabs}
\usepackage{tabularx}
\usepackage{array}
\usepackage{makecell}

\definecolor{acqblue}{RGB}{230,238,250}      
\definecolor{evalgreen}{RGB}{231,243,234}    
\definecolor{controlorange}{RGB}{248,236,221}

\usepackage{microtype}

\usepackage{inconsolata}

\usepackage{graphicx}

\usepackage{array}

\title{From Search to Decision: How Post-Training Shapes Information-Seeking Capabilities in Large Language Models}

\author{
  \textbf{Haohang Huang\textsuperscript{1},
  Xuan Lu\textsuperscript{2}, 
  Mingjin Kuai\textsuperscript{3}, 
  Yuxuan Zhang\textsuperscript{4}, 
  Wei Ji\textsuperscript{3},
  Haiyang Zhang\textsuperscript{1}\thanks{Corresponding author}%
  }
  \\
  \textsuperscript{1}Xi'an Jiaotong-Liverpool University \\
  \textsuperscript{2}Shanghai Jiaotong University \quad
  \textsuperscript{3}Nanjing University \\
  \textsuperscript{4}University of British Columbia \\
  \texttt{haohang.huang25@student.xjtlu.edu.cn}
}

\begin{document}
\maketitle
\begin{abstract}
Information seeking in large language model (LLM) systems increasingly involves decisions about when to acquire information, where to search, and how to evaluate and use evidence. Recent post-training methods have improved several such behaviors, but their objectives and evaluations span different tasks and system settings. This survey organizes research around six interrelated information-seeking capabilities and examines the learning signals, optimization methods, update loci, and credit-assignment mechanisms used to develop them. We synthesize empirical progress in targeted behaviors, end-to-end performance, search control, and generalization. Existing studies report gains in retrieval routing, query refinement, multi-tool exploration, evidence-grounded reasoning, and resource-aware search, although the scope of behavioral evidence varies across capabilities. We identify open challenges in estimating prospective retrieval utility, coordinating information-seeking decisions, and adapting to unfamiliar retrieval environments, and discuss evaluation practices for interpreting future progress.
\end{abstract}

\section{Introduction}
\label{sec:introduction}

Large language models (LLMs) increasingly interact with external information sources to address knowledge-intensive tasks. In such systems, retrieval is no longer necessarily a fixed preprocessing step: models may decide whether to search, formulate queries, explore sources, evaluate and integrate evidence, and revise their next actions in response to observations \citep{asai2024self,jin2025search}. These interrelated decisions constitute information-seeking capabilities that can be shaped through post-training. Recent work has made progress on several of these capabilities. Retrieval-trigger and utility-aware training improve acquisition decisions \citep{labruna2025retrieve,zou2026cobra}; query refinement and multi-tool coordination address information needs and exploration \citep{chen2026qref,luo2025marag}; evidence-grounded methods target support judgments and integration \citep{huang2026rlseek,wang2025improving}; and budget-aware policies improve the efficiency of iterative search \citep{sun2026one}. These advances employ different supervision signals, optimization objectives, and update targets, making their contributions difficult to compare solely through downstream task scores. 
Existing surveys have examined retrieval--reasoning interactions \citep{li2025towards,wei2026survey}, deep-search architectures and workflows \citep{xi2025survey}, and reinforcement learning for agentic search and deep research \citep{lin2025comprehensive,li2025reinforcement}. Related work has also reviewed credit assignment in agentic reinforcement learning \citep{zhang2026reasoning}. However, these surveys primarily organize existing research around system architectures, retrieval processes, or optimization techniques, leaving the learning and empirical progress of individual information-seeking capabilities less systematically examined. To address this gap, we provide a capability-centered synthesis that connects the decisions models learn to make, the post-training mechanisms used to improve them, and the empirical evidence supporting their improvement. This perspective enables a systematic comparison of learning progress across capabilities, evaluation settings, and information environments.

Specifically, we (1) characterize six information-seeking capabilities and their observable behaviors (Section~\ref{sec:what_to_learn}); (2) synthesize the post-training signals, optimizers, update loci, and credit-assignment mechanisms used to improve them (Section~\ref{sec:how-to-learn}); and (3) examine empirical progress and the scope of supporting evidence, including transfer results (Section~\ref{sec:learned}). Following a shared task formulation (Section~\ref{sec:task-formulation}), we conclude by outlining capability-specific evaluation practices and open challenges (Section~\ref{sec:evaluation}).

\section{Task Formulation}
\label{sec:task-formulation}

We consider an LLM-based system that interacts with an external
information environment $\mathcal{E}$ to solve a task $q$.
At step $t$, a decision policy selects an action
$a_t \sim \pi_\theta(\cdot \mid q, h_t)$, where
$h_t = \{(a_i, o_i)\}_{i<t}$ denotes the history of actions
and observations. These actions may involve deciding whether
to acquire information, formulating or refining queries,
exploring information sources, evaluating retrieved evidence,
and continuing or terminating the search. Information-acquisition
actions produce observations $o_t$ from $\mathcal{E}$, which
inform subsequent decisions. The system ultimately generates
an answer $\hat{y} = g(q, h_{T+1})$. This formulation encompasses both single-step retrieval and
iterative information seeking \citep{asai2024self,jin2025search}.
Rather than treating information seeking as a fixed sequence
of pipeline stages, we view it as a collection of interrelated
decision capabilities that operate over evolving information
states. These capabilities may be learned through updates to
an integrated LLM policy or specialized components. We use
this formulation to examine how post-training mechanisms
shape information-seeking decisions and how the resulting
behaviors are evaluated across tasks and environments.

\section{What to Learn: Information-Seeking Capabilities}
\label{sec:what_to_learn}

\begin{figure*}[t]
    \centering
    \includegraphics[width=\textwidth]{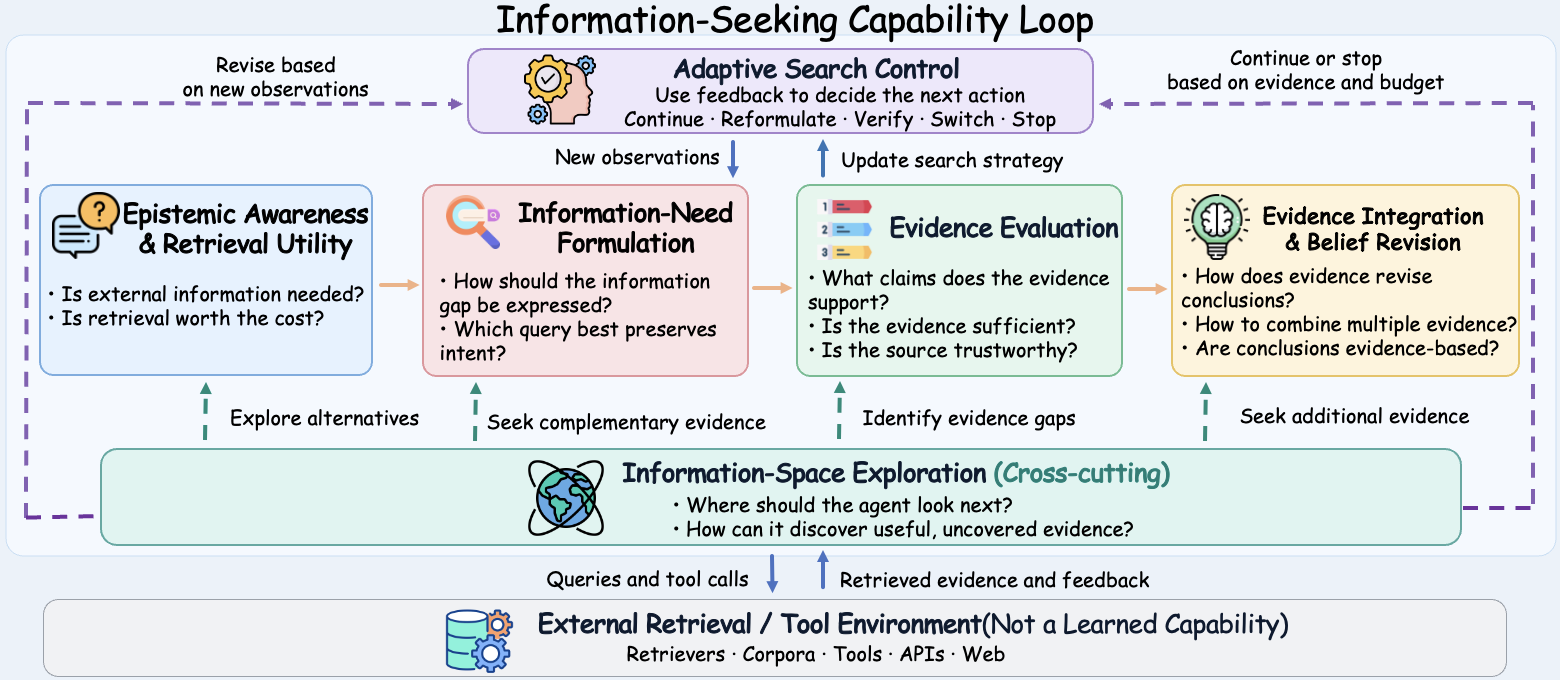}
    \caption{Information-seeking as a capability loop. Six interrelated capabilities support iterative information acquisition, evidence evaluation, and integration.}
    \label{fig:information_seeking_loop}
\end{figure*}

Effective information seeking involves deciding whether to acquire information, expressing what remains unknown, locating useful sources, evaluating and integrating evidence, and revising subsequent actions. We distinguish six interrelated capabilities by the \emph{decisions} they govern, rather than treating them as fixed stages of a retrieval pipeline. Figure~\ref{fig:information_seeking_loop} illustrates their interactions, highlighting the cross-cutting role of exploration and the feedback-driven nature of adaptive search control. These distinctions define the target behaviors of information-seeking policies, while subsequent sections examine how post-training affects these behaviors and what evidence supports their acquisition and generalization.

\subsection{Epistemic Awareness and Retrieval Utility}
\label{sec:epistemic_awareness}

This capability concerns whether a model should acquire external information given its current knowledge and the expected cost of retrieval. Three decisions are related but distinct: recognizing uncertainty, deciding to retrieve, and estimating whether retrieval is worth its cost. \emph{Retrieval necessity} includes learned triggers and knowledge-boundary estimates \cite{labruna2025retrieve,zhang2025kbm}, alongside inference-time uncertainty triggers and assessments of self-knowledge-based routing \cite{su2024dragin,yao2025seakr,moskvoretskii2025adaptive}. In contrast, \emph{retrieval utility} concerns model-specific benefits and acquisition costs \cite{zhang2025llm,wu2026knowledge}. \textit{CoBRA} and \textit{s3} use counterfactual or downstream gains, while other methods employ information-gain signals or internal--external knowledge coordination \cite{zou2026cobra,jiang2025s3,hu2026optimizing,huang2025reinforced}. Cost-aware exploration may also use supplied environmental priors \cite{ding2026calibrate}. The observable target is a retrieval decision that responds appropriately to expected benefit and cost, not merely task difficulty. Realized gains provide useful supervision for retrieval decisions, while calibrated ex-ante utility prediction represents a complementary evaluation target.

\subsection{Information-Need Formulation}
\label{sec:information_need_formulation}
 
\textit{Information-need formulation} concerns how an agent expresses the information missing from its current reasoning state. Its decision target is the content and structure of a query, with the preservation of semantic intent distinct from retrieval success. \emph{Query reformulation} improves an individual request through learned rewriting or retriever feedback \cite{ma2023query,jiang2025deepretrieval}. \emph{Expansion and decomposition} generate complementary queries or parallel subquestions for more complex information needs \cite{zhao2025beyond,zhao2025parallelsearch}. \emph{Reasoning-conditioned refinement} revises queries as new gaps emerge: \textit{StepSearch} optimizes intermediate search behavior, while \textit{QREF} connects refinement with reasoning requirements and evidence coverage \cite{zheng2025stepsearch,chen2026qref}. The relevant behavior is whether a query expresses the unresolved need and recovers necessary evidence, rather than merely obtaining a high retrieval score. \textit{Environment-aware IR} shows that effective query patterns may depend on a particular backend \cite{yuan2026understanding}. Thus, semantic fidelity, retriever-specific effectiveness, and transfer across retrieval environments describe different aspects of formulation.

\subsection{Information-Space Exploration}
\label{sec:information_space_exploration}

Whereas formulation determines \emph{what to ask},
\textit{information-space exploration} concerns \emph{where to
look next} to discover useful and complementary evidence.
\textit{MARAG-R1} learns coordination across retrieval tools,
while diversity-driven policy optimization encourages broader
search trajectories \cite{luo2025marag,chen2027restoring}.
The former emphasizes complementary source coverage, whereas
the latter promotes diversity within exploration. Evaluation studies---\textit{InfoMosaic-Bench},
\textit{Table-as-Search}, \textit{GISA}, and analyses of real
search requests---characterize multi-source seeking, coverage
states, and realistic exploration patterns
\cite{du2026infomosaic,lan2026table,zhu2026gisa,ning2026agentic}.
\textit{SearchOS-V1} further uses explicit system state to
manage coverage and search failures \cite{zhang2026searchos}.
Transferable tool selection and tool discovery extend this
perspective to information-access interfaces
\cite{zhou2026training,dai2026toolsearcher}.
Together, these studies highlight the distinction between
query diversity, evidence coverage, and information gain,
positioning exploration as a cross-cutting capability for
discovering information that addresses unresolved needs.

\subsection{Evidence Evaluation}
\label{sec:evidence_evaluation}

\textit{Evidence evaluation} determines what acquired information warrants concluding. Three judgments are particularly relevant. \emph{Claim support} asks whether a passage supports a specific assertion. \textit{CRAG} assesses retrieval quality, while \textit{RLSeek} and \textit{EAPO} target evidence-related judgments or evidence-conditioned reasoning \cite{yan2024corrective,huang2026rlseek,guan2026evidence}. \textit{RAGTruth} and \textit{RAGChecker} provide complementary diagnoses of unsupported generation and retrieval--generation failures \cite{niu2024ragtruth,ru2024ragchecker}. \emph{Sufficiency} asks whether the accumulated information supports the required inference: \textit{S2G-RAG} models remaining information gaps, while \textit{Sufficient Context} distinguishes missing evidence from failures to use available evidence \cite{li2026s2g,joren2025sufficient}. \emph{Source reliability} asks whether the available support should be trusted, particularly when sources disagree. \textit{AuthorityBench} studies authority perception, and studies of conflicting evidence evaluate responses to incompatible references \cite{yao2026authoritybench,ge2025resolving,yuan2026benchmarking}. \textit{(D)RAGged Into a Conflict} distinguishes conflict types, while counterfactual analyses of convincing evidence examine how source preferences respond to non-factual cues \cite{cattan2025dragged,wan2024evidence}. These research directions target different observable judgments: support, adequacy, and credibility. In particular, detecting agreement with a retrieved passage is not equivalent to establishing the reliability of that passage.

\subsection{Evidence Integration and Belief Revision}
\label{sec:evidence_integration}

\textit{Evidence integration} concerns how acquired information
updates conclusions and supports subsequent reasoning.
It comprises three complementary behaviors: revising beliefs,
composing distributed facts, and faithfully using evidence.

\textbf{Belief revision.}
When retrieved claims conflict with parametric knowledge,
the model must reconcile competing information.
\textit{ClashEval}, \textit{Belief-R}, and source-balancing
studies examine responses to conflicting assertions
\cite{wu2024clasheval,wilie2024belief,li2026large}.
\textit{GuarantRAG} encourages document-grounded answers
through preference learning, while prior-guided reasoning
optimizes the interaction between internal and external
knowledge \cite{zhao2026guaranteeing,yuan2026efficient}.
\textit{FaithfulRAG} addresses fact-level conflicts at
inference time \cite{zhang2025faithfulrag}. Together, these
approaches investigate selective belief revision rather
than unconditional acceptance of external information.

\textbf{Multi-evidence composition.}
Complex questions often require combining complementary facts
across documents. \textit{RAFT} trains evidence-grounded
generation amid distractors, \textit{BRIEF} compresses
retrieved information for multi-hop reasoning, and
\textit{TRACE the Evidence} constructs explicit evidence
chains \cite{zhang2024raft,li2025brief,fang2024trace}. These approaches address complementary aspects of
evidence selection, consolidation, and composition.

\textbf{Faithful evidence utilization.}
\textit{CARE} encourages grounded reasoning, while
\textit{MIRAGE} analyzes passage attribution
\cite{wang2025improving,qi2024model}.
\textit{Counterfactual RAG} probes responses to alternative
evidence \cite{qin2026counterfactual}, while other evaluations
examine context faithfulness and attribution biases
\cite{fadeeva2026faithfulness,abolghasemi2025evaluation}.
These studies distinguish \emph{output reliance}
(answers change with evidence), \emph{compositional reliance}
(necessary facts are combined), and \emph{process-level
reliance} (evidence causally influences decisions),
providing complementary perspectives on faithful
information use.

\textbf{Cross-turn persistence.}
Retaining and updating evidence across interactions extends
integration beyond individual answers. Memory benchmarks
and management methods examine how acquired information
is stored, revised, and reused in subsequent decisions
\cite{wu2024longmemeval,hu2026evaluating,yan2026memory,zhang2026memrl}.

\subsection{Adaptive Search Control}
\label{sec:adaptive_search_control}

\textit{Adaptive search control} concerns the next action after a new observation: continue searching, reformulate, verify, switch sources, or stop. The defining target is an \emph{observation-conditioned decision}, not merely the execution of multiple search steps. \textit{Self-RAG} incorporates retrieval and critique into generation; \textit{Search-R1} and \textit{R1-Searcher} optimize iterative search trajectories; and \textit{ZeroSearch} studies search-policy training with simulated interactions \cite{asai2024self,jin2025search,song2025r1,sun2025zerosearch}. Work on beneficial search behaviors also identifies verification and recovery as relevant actions \cite{jin2026beneficial}. Other approaches focus on \emph{resource allocation}: \textit{AutoSearch} learns search depth, while \textit{One Policy, Any Budget} studies search decisions under varying budgets \cite{sun2026autosearch,sun2026one}. Retrospective critics and contribution-weighted learning target intermediate search actions during training \cite{zhang2026criticsearch,wang2026enhancing}. These directions support different control decisions. Budget adaptation and feedback responsiveness represent complementary dimensions of adaptive search control.

\section{How to Learn: Post-training Mechanisms}
\label{sec:how-to-learn}

The six information-seeking capabilities outlined in Section~\ref{sec:what_to_learn} are targeted by a range of post-training approaches. We examine these approaches along four complementary dimensions: learning signals, policy optimization, update loci, and credit assignment. These dimensions characterize which behaviors are supervised, how feedback is incorporated into learning, which components are updated, and how delayed outcomes are attributed to intermediate decisions. A comparison of representative methods is provided in Appendix~\ref{app:posttraining-mechanisms}.

\subsection{Training Data and Learning Signals}
\label{sec:learning-signals}

\paragraph{Demonstrations and interaction data.}
Demonstrations specify desirable actions or evidence-conditioned responses. \citet{labruna2025retrieve} supervise retrieval triggering, whereas RAFT trains evidence-grounded answering with relevant and distractor documents \citep{zhang2024raft}. Such data provide direct targets for retrieval decisions and evidence integration (Sections~\ref{sec:epistemic_awareness} and~\ref{sec:evidence_integration}), but their coverage is limited to the states represented in the collected examples. Interaction trajectories instead expose models to sequential decisions and observations. ZeroSearch reduces the cost of generating such trajectories through a learned search simulator \citep{sun2025zerosearch}. The trade-off concerns supervision coverage, interaction cost, and the fidelity of training observations.

\paragraph{Rewards and preferences.}
Outcome rewards supervise overall task success without requiring labels for individual search actions, as in Search-R1 \citep{jin2025search}. Behavior-specific signals provide more localized guidance: CoBRA uses paired tool/no-tool outcomes for retrieval decisions \citep{zou2026cobra}; StepSearch combines intermediate search rewards with redundancy penalties \citep{zheng2025stepsearch}; and RLSeek supervises evidence-grounded verification \citep{huang2026rlseek}. Preference-based supervision offers another approach to evidence integration: GuarantRAG uses contrastive DPO to favor document-grounded answers, alongside a separate joint-decoding mechanism \citep{zhao2026guaranteeing}.

\paragraph{Cross-paper comparison.}
These signals differ in the behavioral distinctions they expose to optimization. Search-R1 uses terminal correctness to jointly optimize search and reasoning, whereas CoBRA uses counterfactual reward margins to target whether external information should be acquired. For query formulation, StepSearch combines search-key alignment, information gain, and answer-level feedback across intermediate steps \citep{zheng2025stepsearch}; QREF instead compares original and refined query batches through entity-level evidence recovery and preservation \citep{chen2026qref}. The former emphasizes step-wise search progression, while the latter directly rewards changes in retrieval coverage. Thus, learning signals differ not only in granularity but also in the decisions they distinguish, ranging from acquisition boundaries and query refinement to overall task completion.

\subsection{Policy Optimization}
\label{sec:policy-optimization}

Optimization methods differ in how they use supervision and obtain training states. Supervised fine-tuning (SFT) learns from demonstrated actions without policy rollouts, making it suitable for well-specified behaviors but dependent on demonstration coverage. Preference optimization learns relative preferences among candidate outputs, as in GuarantRAG's evidence-conditioned training \citep{zhao2026guaranteeing}. Reinforcement learning (RL) optimizes policies through sampled interactions: Search-R1 trains interleaved reasoning and retrieval \citep{jin2025search}, while AnySearch incorporates accuracy--efficiency feedback across search budgets \citep{sun2026one}. This supports optimization over sequential decisions but introduces rollout and environment-interaction costs. Hybrid approaches combine supervised initialization with interactive refinement, as in CoBRA's retrieval-boundary learning and MARAG-R1's multi-retriever coordination \citep{zou2026cobra,luo2025marag}. Teacher-guided optimization provides an additional mechanism. SSPO uses teacher--student disagreement to weight step-level advantages within GRPO, while outcome rewards determine the direction of policy updates \citep{wu2026beyond}. Teacher information therefore modulates reward-based learning rather than replacing it with direct imitation. Across these approaches, the optimizer determines how supervision updates the policy, while the training signal determines which behaviors receive feedback. Their interaction also depends on the training environment: controlled comparisons show that corpus coverage and rollout protocols can substantially affect results under a shared optimizer \citep{zhao2026retrieval}.

\subsection{Learning Targets and Update Loci}
\label{sec:update-loci}

The update locus determines whether training specializes a component or adapts a broader policy. QREF trains a query refiner without retraining the underlying reasoning agent, focusing on information-need formulation (Section~\ref{sec:information_need_formulation}) \citep{chen2026qref}. Search-R1 instead updates a policy that jointly produces reasoning and search actions \citep{jin2025search}. Specialized components also target evidence processing: BRIEF learns multi-document compression, while RLSeek optimizes evidence-grounded verification \citep{li2025brief,huang2026rlseek}. Modular updates concentrate supervision on particular operations and preserve other components, whereas integrated policies allow coupled decisions to adapt jointly. A further distinction concerns parameter updates versus external state. Memory-R1 trains memory-management and utilization policies, while MemRL updates utility estimates for external episodic memory with frozen language-model weights \citep{yan2026memory,zhang2026memrl}. Agentic-R iteratively co-optimizes the retriever and agent \citep{liu2026agentic}. Joint optimization can accommodate interactions between retrieval and reasoning, although it requires coordinating updates across components. These approaches illustrate complementary routes to improving information-seeking capabilities, differing in adaptation scope, modularity, and component coupling.

\subsection{Long-horizon Credit Assignment}
\label{sec:credit-assignment}

In multi-turn search, final-answer rewards provide task-level supervision but limited attribution to individual actions. Intermediate signals address this sparsity: StepSearch uses information-gain and redundancy rewards, while AutoSearch derives depth-related supervision from intermediate answers \citep{zheng2025stepsearch,sun2026autosearch}. Retrospective methods instead use completed trajectories or gold-conditioned estimates to provide turn-level feedback. CriticSearch employs a retrospective critic to assess individual search turns \citep{zhang2026criticsearch}, while CW-GRPO reweights outcome-based advantages according to judged round contributions \citep{wang2026enhancing}. TRACE and SSPO further illustrate different constructions of intermediate credit \citep{tao2026trace,wu2026beyond}.

\paragraph{Credit construction and behavioral targets.}
TRACE estimates turn-level contributions through changes in gold-answer-conditioned state values computed by a frozen reference model \citep{tao2026trace}. SSPO instead uses teacher--student disagreement to adjust step-level advantage weights within GRPO \citep{wu2026beyond}. The former distributes feedback according to estimated progress between search states, whereas the latter prioritizes updates at steps where teacher guidance and student behavior diverge. This distinction concerns the basis of credit assignment, beyond its temporal granularity. Intermediate progress signals can guide query formulation and search control, while retrospective estimation provides feedback across longer trajectories. These mechanisms also impose different requirements: TRACE uses gold-answer-conditioned reference-model evaluations, whereas SSPO requires teacher-side evidence and additional forward computation. Across these approaches, credit assignment determines how delayed outcomes influence intermediate decisions, complementing the behavioral targets specified by learning signals and the components selected for optimization. Their relative effectiveness depends on the targeted decisions, trajectory structure, and available supervision. The empirical implications for capability improvement and generalization are examined in the following section.

\section{What Was Actually Learned?}
\label{sec:learned}

Recent work provides growing evidence that post-training improves information-seeking behaviors across the six capabilities. To understand what these improvements establish, we synthesize empirical progress along three complementary dimensions: measurement alignment, which connects reported metrics to target behaviors; attribution, which examines the contributions of trained decisions and components; and generalization, which characterizes the conditions under which improvements persist. We first examine evidence for targeted behavioral learning (Section~\ref{sec:alignment}), then consider end-to-end improvements and intermediate credit assignment (Sections~\ref{sec:attribution}--\ref{sec:hindsight}), and finally assess generalization across information environments (Section~\ref{sec:transfer}). Together, these dimensions provide a basis for interpreting capability learning across heterogeneous evaluation settings. Table~\ref{tab:empirical_evidence} summarizes representative results, while Appendix~\ref{app:capability_evaluation} provides complementary evaluation designs.

\begin{table*}[t]
\centering
\footnotesize
\setlength{\tabcolsep}{3pt}
\renewcommand{\arraystretch}{1.08}

\begin{tabularx}{\textwidth}{@{}
>{\raggedright\arraybackslash}p{0.15\textwidth}
>{\raggedright\arraybackslash}X
>{\raggedright\arraybackslash}p{0.19\textwidth}
>{\raggedright\arraybackslash}p{0.16\textwidth}@{}}

\toprule
\textbf{Capability} &
\textbf{Progress \& measure} &
\textbf{Evidence boundary} &
\textbf{Evaluated settings} \\
\midrule

\rowcolor{acqblue}
Epistemic awareness &
Retrieval routing and utility
\citep{labruna2025retrieve,jiang2025s3,zou2026cobra,ding2026calibrate};
accuracy and cost. &
Calibrated ex-ante utility. &
OOD tool-use tasks. \\
\midrule

\rowcolor{acqblue}
Information-need formulation &
Query rewriting and decomposition
\citep{ma2023query,jiang2025deepretrieval,zhao2025parallelsearch,chen2026qref};
retrieval gain and coverage. &
Semantic intent preservation. &
Different agent backbones. \\
\midrule

\rowcolor{acqblue}
Information-space exploration &
Multi-tool and diverse search
\citep{luo2025marag,chen2027restoring};
recall and diversity. &
Discovery of necessary evidence. &
Multiple tasks; fixed tools. \\
\midrule

\rowcolor{evalgreen}
Evidence evaluation &
Support and sufficiency assessment
\citep{li2026s2g,huang2026rlseek,guan2026evidence};
span and evidence metrics. &
Source reliability judgment. &
OOD detection tasks. \\
\midrule

\rowcolor{evalgreen}
Evidence integration &
Grounded and multi-hop reasoning
\citep{zhang2024raft,li2025brief,wang2025improving,zhao2026guaranteeing};
QA and hallucination metrics. &
Causal evidence reliance. &
Real and counterfactual QA. \\
\midrule

\rowcolor{controlorange}
Adaptive search control &
Iterative and budget-aware search
\citep{jin2025search,sun2026autosearch,sun2026one,zhang2026criticsearch};
accuracy and efficiency. &
Feedback-sensitive next actions. &
Unseen search budgets. \\

\bottomrule
\end{tabularx}

\caption{Empirical evidence across six information-seeking capabilities. Evidence boundaries refer to the cited evaluations.}
\label{tab:empirical_evidence}
\end{table*}

\subsection{Evidence for Targeted Behavioral Learning}
\label{sec:alignment}

Across the six capabilities, post-training has produced improvements measured through different aspects of information-seeking behavior. Retrieval routing and budget-aware control are evaluated through relatively direct action-level measures, including tool-use decisions and resource allocation \citep{zou2026cobra,sun2026one}. Query refinement is assessed through retrieval effectiveness and evidence coverage, capturing how well generated queries recover useful information under a given retrieval setup \citep{chen2026qref}. Multi-tool and diversity-oriented search further extend evaluation toward information-space coverage \citep{luo2025marag,chen2027restoring}. These results collectively demonstrate progress in deciding when, how, and where to acquire information, while distinguishing retrieval utility, query effectiveness, and exploration coverage as separate behavioral targets. Evidence-related capabilities exhibit a complementary pattern. Span-level support detection and sufficiency assessment provide localized measurements of evidence evaluation, whereas grounded-answer accuracy and counterfactual performance capture the downstream effects of evidence integration \citep{huang2026rlseek,li2026s2g,wang2025improving}. Comparing these studies reveals a spectrum of evidence, from observable actions and retrieval outcomes to evidence judgments and context-conditioned generation. Each provides useful information about learned behavior, but supports a different level of interpretation. In particular, retrieval gains characterize query effectiveness more directly than semantic fidelity, while answer-level improvements provide a broader measure of evidence use than process-level causal reliance. The central finding is therefore that behavioral learning is increasingly measurable across capabilities, with the correspondence between measurement and target behavior determining the specificity of the supported conclusion.

\subsection{Interpreting End-to-End Improvements}
\label{sec:attribution}

End-to-end approaches provide substantial evidence of improved information-seeking performance. Search-R1 demonstrates the effectiveness of outcome-rewarded iterative search, while Agentic-R reports further gains through joint retriever--agent optimization \citep{jin2025search,liu2026agentic}. Together, these studies show that post-training can improve the coordination of search and reasoning, whether through policy optimization or coupled component updates. Cross-paper comparisons help explain the sources of these improvements. Under controlled training conditions, \citet{zhao2026retrieval} find that corpus coverage substantially affects performance and method comparisons, with outcome-based rewards remaining competitive against more elaborate credit assignments. Component-level analyses provide complementary evidence: QREF examines the respective contributions of query refinement and evidence filtering, while controlled changes to user and document assertions reveal how models adjust their reliance on external information \citep{chen2026qref,li2026large}. Together, these results show that end-to-end improvements can reflect changes in search policies, retrieval infrastructure, and evidence utilization. Component ablations and controlled comparisons help attribute observed gains to these different sources, complementing the behavior-specific measurements discussed in Section~\ref{sec:alignment}.

\subsection{Credit Assignment and Feedback-Conditioned Behavior}
\label{sec:hindsight}

Recent studies demonstrate several effective ways to exploit intermediate supervision for long-horizon search. AutoSearch improves search-depth efficiency through intermediate answer evaluation, while CriticSearch uses retrospective assessments to improve task performance and training convergence \citep{sun2026autosearch,zhang2026criticsearch}. TRACE similarly reports improvements using state-value-based turn-level credit assignment \citep{tao2026trace}. SSPO further shows that step-level weighting can mitigate premature termination associated with distillation from a teacher with privileged evidence \citep{wu2026beyond}. Collectively, these results show that intermediate supervision can improve search efficiency, learning stability, and long-horizon task performance. A cross-paper comparison reveals distinct sources of training feedback: intermediate answer quality, retrospective trajectory assessment, state-value estimation, and teacher-based signals. Their reported gains span training convergence, task accuracy, and search efficiency, providing complementary evidence for effective long-horizon policy learning. At deployment, search depth and budget use characterize resource allocation, while feedback-conditioned evaluations examine how a policy adapts its next action to new retrieval observations. This distinction extends the attribution analysis by connecting the distribution of training feedback to the behaviors demonstrated by the learned policy.

\subsection{Generalization Across Information Environments}
\label{sec:transfer}

Existing studies report encouraging generalization results across several dimensions. AnySearch demonstrates improvements under unseen budget constraints, while CoBRA evaluates tool-dependent out-of-distribution questions \citep{sun2026one,zou2026cobra}. QREF reports improvements across agent backbones under a shared web-search pipeline, providing evidence of transfer across model configurations \citep{chen2026qref}. Evidence-related studies examine other variations: RLSeek reports out-of-domain hallucination detection results, and CARE evaluates evidence utilization on real-world and counterfactual QA tasks \citep{huang2026rlseek,wang2025improving}. Memory-R1 further reports improvements across memory benchmarks and model scales \citep{yan2026memory}. Together, these results indicate that learned information-seeking behaviors can remain effective under selected changes in tasks, models, and resource constraints. The interpretation of these findings depends on which environmental conditions change and which remain fixed. Budget generalization evaluates resource adaptation, whereas backbone transfer examines performance across model configurations under a shared retrieval setting. TRACE extends evaluation from closed-retrieval training to open-web benchmarks, reporting cross-environment outcome improvements \citep{tao2026trace}. Retriever-aware query learning, meanwhile, identifies backend-specific query preferences and reduced effectiveness when learned strategies are applied to different retrievers \citep{yuan2026understanding}. These findings highlight both the benefits of environment-specific adaptation and the importance of evaluating transfer across distinct retrieval conditions. Taken together, the literature establishes measurable progress in information-seeking behaviors across all six capabilities. Our synthesis reveals three complementary aspects of this progress: capability-specific metrics identify which behaviors improve, end-to-end and component-level analyses clarify the sources of these improvements, and generalization studies characterize the conditions under which learned behaviors remain effective. These findings provide a differentiated account of what post-training has taught LLMs, connecting observable behavioral gains to their learning mechanisms and evaluation conditions.

\section{Evaluation Principles and Open Problems}
\label{sec:evaluation}

\subsection{Capability-Specific Evaluation}
\label{sec:capability-evaluation}

The findings in Section~\ref{sec:learned} suggest three practical evaluation principles. First, \emph{capability-specific measures} should complement end-task performance with observable behaviors, such as retrieval decisions, evidence coverage, support judgments, or search efficiency \citep{zou2026cobra,chen2026qref,huang2026rlseek,sun2026one}. Second, \emph{controlled evaluation} should isolate the targeted decision while specifying the updated component and information available to the deployed policy. For example, a \emph{fixed-history feedback intervention} holds the task, preceding trajectory, and remaining budget constant, varies only the latest retrieval observation, and measures whether the next action appropriately continues, reformulates, verifies, or stops. Similar interventions can vary query content or necessary evidence while controlling the retriever and downstream components. Third, \emph{explicit generalization settings} should identify the changed conditions, such as retrievers, corpora, tools, or budgets, and report the corresponding behavioral outcomes.

\paragraph{Capability evaluation protocol.}
For consistent reporting, we recommend specifying the target behavior, learning intervention and updated components, behavioral measures, controlled factors, generalization setting, and supported conclusions, including untested claims. This checklist guides interpretation without requiring every study to evaluate all possible conditions. Appendix~\ref{app:capability_evaluation} provides capability-specific designs and potential confounders.

\subsection{Open Problems}
\label{sec:open-problems}

Three challenges emerge from the preceding analysis. \emph{First, prospective acquisition learning.} Counterfactual and cost-aware training have improved retrieval decisions \citep{zou2026cobra,ding2026calibrate}, but learning to anticipate the value of unavailable evidence under changing costs and source quality remains challenging. Training across diverse acquisition conditions and evaluating pre-retrieval decisions offer complementary directions. \emph{Second, coordinated capability learning.} Query refinement, multi-tool exploration, and evidence-grounded reasoning have shown individual improvements \citep{chen2026qref,luo2025marag,wang2025improving}. Learning to coordinate these decisions under incomplete or conflicting information may benefit from supervision linking information gaps, evidence coverage, and subsequent use. Its contribution should be examined against outcome-only training and component-level controls \citep{zhao2026retrieval}. \emph{Third, cross-environment adaptive control.} Budget-aware training and finer-grained credit assignment improve long-horizon search \citep{sun2026one,tao2026trace}, while adapting to unfamiliar retrieval feedback remains a distinct learning challenge. Training across heterogeneous search environments, combined with controlled feedback interventions, could help assess whether adaptive behavior extends beyond familiar conditions.

\section{Conclusion}
\label{sec:conclusion}

Post-training improves retrieval decisions, query refinement, evidence use, and adaptive search, but the strength and generality of supporting evidence differ across capabilities. Our capability-centered framework links target decisions, learning mechanisms, behavioral measurements, and evaluation conditions to clarify what reported results establish and what remains untested. Future research should prioritize prospective retrieval utility, coordination across capabilities, and adaptation to unfamiliar information environments.



\bibliography{custom}

\appendix


\section{Post-training Mechanism Comparison}
\label{app:posttraining-mechanisms}

This appendix provides a detailed comparison of representative post-training approaches discussed in Section~\ref{sec:how-to-learn}. Table~\ref{tab:posttraining-mechanisms} organizes these methods by their primary information-seeking capability, learning signal and credit assignment, and optimization method and update locus. The comparison highlights different ways of learning information-seeking behaviors, including demonstration-based supervision, outcome-level rewards, behavior-specific feedback, and preference optimization. It also distinguishes updates to integrated search policies from those targeting specialized components, such as query refiners and evidence verifiers. Capability assignments reflect the primary training targets rather than mutually exclusive categories, as individual methods may influence multiple related decisions. The table characterizes training designs rather than ranking their effectiveness; empirical improvements and evaluation scope are discussed separately in Section~\ref{sec:learned}.

\begin{table*}[t]
\centering
\footnotesize
\setlength{\tabcolsep}{3pt}
\renewcommand{\arraystretch}{0.95}
\begin{tabularx}{\textwidth}{@{}p{0.22\textwidth}p{0.17\textwidth}X p{0.24\textwidth}@{}}
\toprule
\textbf{Work} &
\textbf{Primary capability} &
\textbf{Learning signal / credit} &
\textbf{Optimization / update locus} \\
\midrule

CoBRA \citep{zou2026cobra} &
Retrieval utility &
Tool/no-tool margins &
SFT + RL / policy \\

QREF \citep{chen2026qref} &
Query formulation &
Demonstrations; entity gain &
SFT + PPO / refiner \\

MARAG-R1 \citep{luo2025marag} &
Exploration &
Demonstrations; coverage rewards &
SFT + RLOO / agent \\

RLSeek \citep{huang2026rlseek} &
Evidence evaluation &
Evidence-grounded verification &
RL / verifier \\

RAFT \citep{zhang2024raft} &
Evidence integration &
Gold/distractor documents &
SFT / generator \\

GuarantRAG \citep{zhao2026guaranteeing} &
Evidence integration &
Answer-pair preferences &
DPO / generator \\

Search-R1 \citep{jin2025search} &
Adaptive control &
Answer correctness (episode) &
PPO/GRPO / policy \\

AutoSearch \citep{sun2026autosearch} &
Adaptive control &
Intermediate answers (depth) &
RL / policy \\

TRACE \citep{tao2026trace} &
Adaptive control &
State-value differences (turn) &
GRPO / policy \\

\bottomrule
\end{tabularx}
\caption{Representative post-training approaches to information-seeking capabilities, compared by primary capability, learning signal, optimization method, and update locus. Capability assignments indicate primary training targets rather than exclusive categories.}
\label{tab:posttraining-mechanisms}
\end{table*}

\section{Capability-Specific Evaluation Resources}
\label{app:capability_evaluation}

Capability-specific evaluation can combine existing benchmarks, method-specific behavioral measures, and controlled interventions. Table~\ref{tab:appendix_capability_evaluation} organizes representative resources and illustrates how the six capabilities can be assessed through target behaviors, observable measures, interventions, potential confounders, and transfer conditions. The illustrative protocols extend the cited evaluation resources with complementary experimental designs.

\begin{table*}[t]
\centering
\footnotesize
\setlength{\tabcolsep}{3pt}
\renewcommand{\arraystretch}{1.15}

\begin{tabularx}{\textwidth}{@{}
>{\raggedright\arraybackslash}p{0.155\textwidth}
>{\raggedright\arraybackslash}X
>{\raggedright\arraybackslash}X
>{\raggedright\arraybackslash}p{0.16\textwidth}
>{\raggedright\arraybackslash}p{0.135\textwidth}@{}}

\toprule
\textbf{Capability / Target behavior} &
\textbf{Measures and resources} &
\textbf{Illustrative intervention} &
\textbf{Potential confounders} &
\textbf{Transfer conditions} \\
\midrule

\textbf{Epistemic awareness and retrieval utility}

Decide whether acquisition is worthwhile. &
Retrieval decisions, paired tool/no-tool gains, and acquisition costs
\citep{labruna2025retrieve,zou2026cobra,ding2026calibrate}. &
Vary retrieval costs or source-quality information; compare decisions made before retrieval with paired acquisition outcomes. &
Prior knowledge, question difficulty, evidence availability. &
Unseen costs, domains, and source distributions. \\

\midrule

\textbf{Information-need formulation}

Express unresolved information needs. &
Query effectiveness, evidence-anchor coverage, and semantic fidelity
\citep{ma2023query,chen2026qref,yuan2026understanding}. &
Substitute candidate queries while holding the corpus, retriever, and reader fixed. &
Retriever-specific query patterns, query length, downstream reasoning. &
Unseen retrievers, domains, and agent backbones. \\

\midrule

\textbf{Information-space exploration}

Discover useful, previously uncovered evidence. &
Multi-source coverage, necessary-evidence recall, and search diversity
\citep{du2026infomosaic,lan2026table,luo2025marag}. &
Remove an informative source while retaining an alternative route to the required evidence. &
Source overlap, tool availability, external coverage scaffolds. &
Unseen sources, tool sets, and information spaces. \\

\midrule

\textbf{Evidence evaluation}

Judge support, sufficiency, and reliability. &
Claim-level support, context sufficiency, and authority judgments
\citep{ru2024ragchecker,li2026s2g,yao2026authoritybench}. &
Independently vary factual support and source-authority cues; introduce conflicting claims. &
Source popularity, surface cues, verifier proxies. &
New domains, source types, and conflict distributions. \\

\midrule

\textbf{Evidence integration}

Revise conclusions and combine necessary facts. &
Grounded-answer quality, belief revision, and cross-turn information use
\citep{wu2024clasheval,wang2025improving,hu2026evaluating}. &
Delete or replace necessary evidence; introduce updated facts across turns and examine subsequent conclusions. &
Parametric knowledge, context position, evidence redundancy, external memory. &
New document sets, evidence conflicts, and temporal changes. \\

\midrule

\textbf{Adaptive search control}

Select subsequent actions from search feedback. &
Search depth, stopping behavior, budget efficiency, and turn-level search outcomes
\citep{sun2026autosearch,sun2026one,zhang2026criticsearch}. &
Hold the task and trajectory prefix fixed while changing the latest retrieval observation. &
Budget instructions, fixed search scaffolds, trajectory differences. &
Unseen budgets, retrieval backends, and feedback conditions. \\

\bottomrule
\end{tabularx}

\caption{Capability-specific evaluation resources and illustrative protocols. Cited studies provide empirical reference points, while the interventions, potential confounders, and transfer conditions outline complementary evaluation designs.}
\label{tab:appendix_capability_evaluation}
\end{table*}

\paragraph{Interpreting capability evidence.}
Behavioral measures and controlled interventions capture complementary aspects of capability learning. For instance, evidence recovery measures query effectiveness, while semantic fidelity can be assessed through preservation of unresolved information needs. Similarly, evidence replacement tests capture answer sensitivity, while process-level analyses examine how evidence influences intermediate decisions. Reporting observed behaviors, controlled factors, and evaluation conditions clarifies the scope of the resulting conclusions.

\paragraph{Generalization settings.}
Transfer can be evaluated across changes in tasks, retrievers, corpora, tools, agent backbones, budgets, and feedback distributions. These conditions capture different aspects of generalization and may vary simultaneously in open environments. Explicitly reporting changed and controlled factors helps characterize the scope of demonstrated transfer and supports comparisons across evaluation settings.

\end{document}